\documentstyle[aps,epsfig,prl]{revtex}

\begin{document}

\title{
 NEGATIVE  SPECIFIC  HEAT   IN  OUT-OF-EQUILIBRIUM NONEXTENSIVE
   SYSTEMS       
}

\author{ A. Rapisarda\textsuperscript{*    }
and     V.  Latora
}

\address{    Dipartimento  di  Fisica  e  Astronomia,   Universit\`{a}
 di Catania  and \\
    INFN  Sezione di Catania, Corso Italia 57,  I-95129  Catania,
  Italy\\---------------------------\\
Talk presented at the International Workshop on Multifragmentation \\
INFN-LNS, Catania 28$^{th}$ November - 1$^{st}$ December 2001\\
$^*$ Corresponding     author,      e-mail:     andrea.rapisarda@ct.infn.it
}
\date{\today}
\maketitle

\begin{abstract}
We discuss the occurrence of  negative  specific  heat in a nonextensive
   system which has an equilibrium  second-order phase transition.
  The specific  heat  is   negative only  in  a  transient regime
 before equilibration ,   in correspondence  to  long-lasting
  metastable states. For these states  standard  equilibrium
   Bolzmann-Gibbs   thermodynamics  does   not apply  and  the
 system  shows   non-Gaussian   velocity distributions, anomalous
 diffusion and correlation in phase space.    Similar   results
   have   recently   been   found  also   in  several    other
   nonextensive   systems,  supporting   the  general   validity
  of  this  scenario. These models seem also to  support  the
 conjecture  that  a nonexstensive statistical formalism, like
 the one  proposed  by  Tsallis,   should be  applied  in  such
 cases.  The   theoretical   scenario  is   not completely clear
  yet , but   there are already many  strong  theoretical  indications
  suggesting  that,  it   can   be  wrong   to  consider    the
   observation  of  an experimental  negative specific heat 
  as  an  unique  and   unambiguous  signature   of   a    standard
   equilibrium  first-order phase transition.
\end{abstract}

\section{
 INTRODUCTION
}

In the last years there has been an increasing interest in phase
 transitions occurring in systems of finite size. Nuclear multifragmentation
 phase transition  is only one of the most interesting examples
 [1]. There has also been a lively debate in the literature on
 how to detect unambiguously the occurrence of a phase transition
 in small systems.  An interesting  proposal has been the measurement
 of  negative specific heat [1,2]. Recently   several experiments
  have found  this signal not only in nuclei  [3 ],  but also
 in atomic clusters [4]. The point we want  to stress here is
 that in standard Bolzmann-Gibbs (BG) thermodynamics, which is
 based on extensive systems, i.e. systems for which energy and
 the entropy are  proportional to N, the specific heat can never
 be negative [5].  However several theoretical pioneering investigations
 have demonstrated that such a property is violated, 
\textit{in the microcanonical}
   ensemble, for nonextensive 
systems [1,6,7],   i.e.  for long-range interactions
 (Coulomb and gravitational forces), but also for finite-size
 systems with short-range interaction. From these studies,  the
 application of BG statistics seems to remain valid for some
 nonextensive systems, if  the microcanonical ensemble is used,
 although  an inequivalence of ensembles remains also in the
 thermodynamic limit [6]. A different approach to tackle nonextensive
 systems has been proposed by Tsallis in 1988 [8] .  Tsallis
 generalized statistics, which contains the BG formalism as a
 particular case, has found since then an increasing and successful
 amount of applications in several fields. The main validity
 of this approach has been found for physical situations where
 out-of-equilibrium phenomena,  long-range corre-lations, long-term
 memory effects  and  anomalous fluctuations are observed [9].
  Recently the appearance of a negative specific heat has  been
 numerically observed  in correspondence to a metastable long-lasting
 regime, where Tsallis formalism and not  the BG one seems to
 apply [10-14].
\textit{Hot nuclear compound systems, }formed in high energy heavy-ion 
reactions\textit{, are nonextensive systems, }therefore\textit{ }
the above mentioned results suggest that  one should be very
 careful in applying  standard equilibrium BG thermodynamics
 in such a  fast  dynamical   phenomenon like multi-fragmentation.
  Even  a small deviation from equilibrium could prevent the
 application of  the BG formalism. In this short contribution,
 we briefly discuss some recent numerical simulations of a nonextensive
  system, the Hamiltonian Mean Field (HMF) model, which illustrate
 the occurrence  of a negative specific only in  a transient
 out-of-equilibrium regime [10]. This model is not a peculiar
 exceptional example and  its anomalous behavior should be  a
 strong warning against a simple and straightforward extrapolation
 of an equilibrium phase transition signature, in finite nonextensive
 systems, when measuring a negative specific heat. The paper
 is organized as follows: we introduce the model in section II,
 numerical simulations are presented in section III and conclusion
 are drawn in section IV.
\newpage
\section{
  THE   MODEL  HAMILTONIAN}

Our arguments are based on the following  system of N fully-coupled
 classical rotators, whose Hamiltonian is given by [15] 

\begin{equation}
H=K+V= \sum_{i=1}^N  {{p_i}^2 \over 2} +
  {1\over{2N}} \sum_{i,j=1}^N  [1-cos(\theta_i -\theta_j)]~~,
\end{equation}

\noindent
where $\theta_i$   is  the
the    angle and  $p_i$  is   the corresponding    conjugate   momentum.
   The canonical analytical   solution   of   the   model predicts
 a  second-order phase transition, whose order parameter   is
   the    total  magnetization M, given by the averaged sum over
 the spin vectors  ${\bf m}_i=(cos \theta_i, sin \theta_i)$ , i.e.

\begin{equation}
{\bf M}={1\over N} \sum_{i=1}^N {\bf m}_i  ~~.
\end{equation}

\noindent
For small energies the absolute value of the magnetization is
 $M \sim 1$  , while for energies greater than the critical value $T_c=0.5
 ~ ~ ~ (U_c=E_c/N=0.75)$
 , we have $M \sim 0$ . The equilibrium caloric curve has been derived
 in ref [15] and  is given by the expression for the energy density

\begin{equation}
U = E/N={T \over 2} + {1\over 2} \left( 1 - M^2 \right) ~,
\end{equation} 

\noindent
T being the temperature. So it is very interesting to compare
 the exact canonical solution (3) with numerical microcanonical
 simulations. Moreover, due to the long-range nature of the interaction,
 this system is nonextensive.  Thus the  model is of  extreme
 interest also from a fundamental point of view, for statistical
 mechanics, whose standard Bolzmann-Gibbs text-books formulation
 is based on the extensive and thus additive properties of  energy
 and entropy.  For these reasons the model   has been intensively
 and extensively  studied in the last years [10-11,15,16].  It
 has also been  slightly modified  to understand   the influence
 of  the range of the interaction [17]. Finite-size effects,
 chaotic dynamics and superdiffusion have been investigated in
 detail  for the HMF model [10-11,15,16].  But an  important
 point which has also been  particularly studied is the relaxation
 to equilibrium: in fact, when out-of-equilibrium initial conditions
 are considered, the model presents a very slow and anomalous
 dynamical behavior, in a energy range below the critical point.
    In the following we will focus our attention    mainly to
 this out-of-equilibrium regime before relaxation, discussing
 its eventual  relevance for nuclear multifragmentation experiments.

\section{
  NUMERICAL  SIMULATIONS
}

We present in this section some numerical simulations which show
 the interesting transient dynamical behavior of the HMF model.
In fig.1a we plot the equilibrium caloric curve  (3) (full line)
 in correspondence of  two numerical results, for N=10000 and
 100000 (open circles and squares), before complete equilibration,
 corresponding  to  a  time  t=1200. The time step used was 0.2.
 The points which mostly disagree with the equilibrium curve
 are in the energy range $0.5\le U \le 0.75$, 
below  the critical energy density,
 indicated as a dashed vertical line. In this region, a complete
 equilibration is generally obtained only after
 10\textsuperscript{6}-10\textsuperscript{7}
 time-steps, according to the size of the system, see ref.[16]
 for more technical details about the integration scheme.

In order to study the dynamics of this slow relaxation, we fix
 a particular energy density, i.e. U=0.69,  and we plot in fig.1b
 the quantity $2<K>/N$ as a function of time, $<K>$  being the average
 kinetic energy. The simulations display  a plateau  for a long
 transient time which does not correspond to the equilibrium
 value $T_{eq}=0.476$  , also reported as a dashed red line. The system
 is trapped in a quasi-stationary state (QSS), whose   whose
 lifetime  increases with N [10]. The quantity  $2<K>/N$ coincides
 with the temperature if a stationary situation exists, thus
 we can refer to the plateau values  as the N-dependent temperatures
 of the quasi-stationary states (QSS). The relaxation is reached,
 as the plot shows,  only after a long time, which increases
 linearly with N. Also the QSS temperature depends on the size
 and converge to the infinite size value $T_{\infty}=0.38$  as a power-law[10].

\begin{figure}
\begin{center}
\epsfig{figure=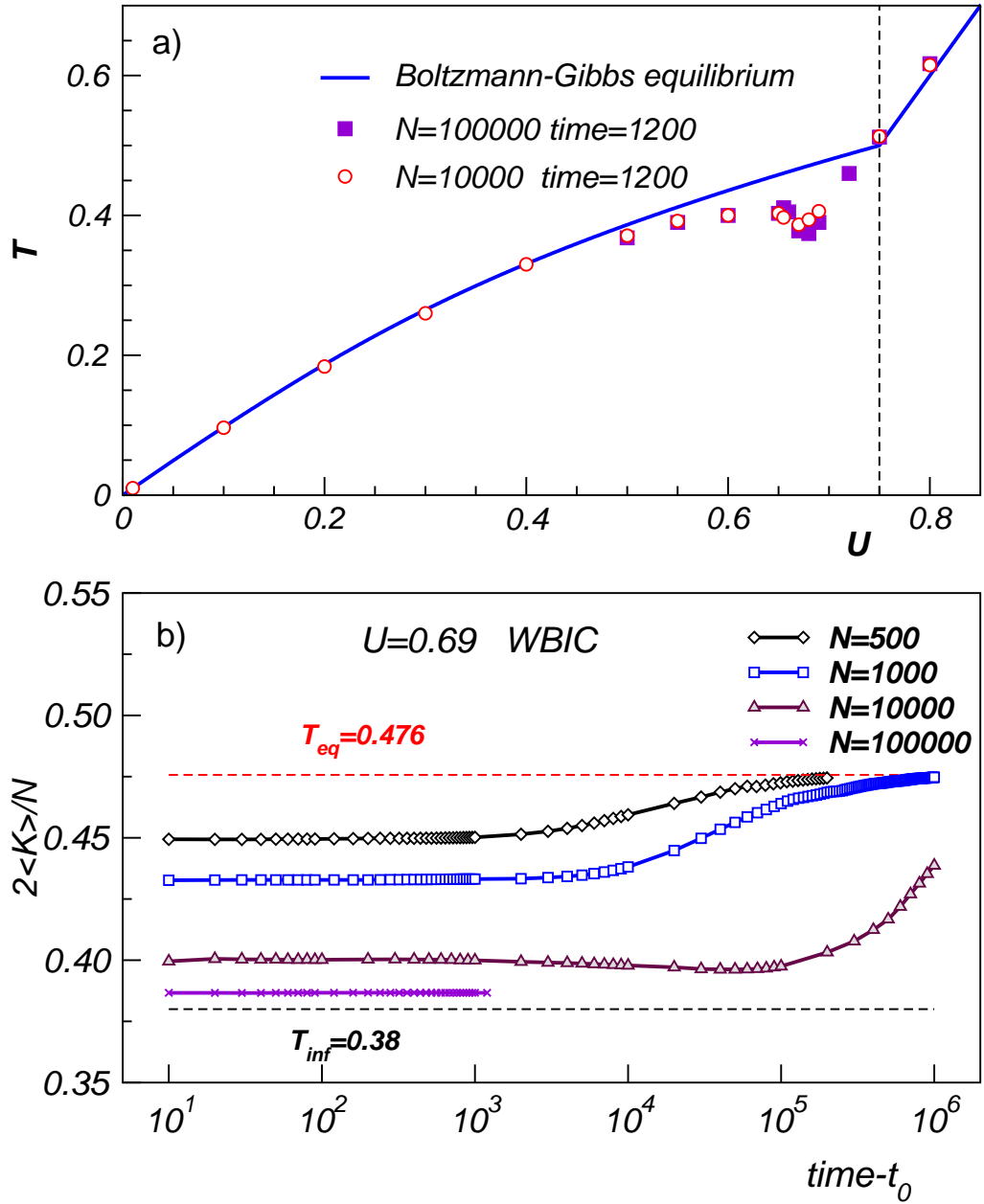,width=17truecm,angle=-90}
\end{center}
\caption{(a) We show the temperature T, calculated by taking
 the average kinetic energy per particle $2<K>/N$ , and the energy
 density $U=0.69$  for HMF systems of different sizes N=10000 (open
  circles) and N=100000 (squares). For comparison  the equilibrium
 caloric curve is also shown as full line. The  numerical simulations
 were initialized considering water bag initial conditions. The
 points are taken after  a short time, t=1200, see text. The
 dashed line indicates the critical point.  (b) We show the temperature
 time evolution   for different N values. The initial part has
 been subtracted $(t_0=100)$. We report also  the equilibrium temperature
 $T_{eq}$  (upper red dashed line) and $T_{\infty}$ the temperature to which
 the QSS tend for  infinite size  (bottom  black dashed line).
}
\end{figure}

\begin{figure}
\begin{center}
\epsfig{figure=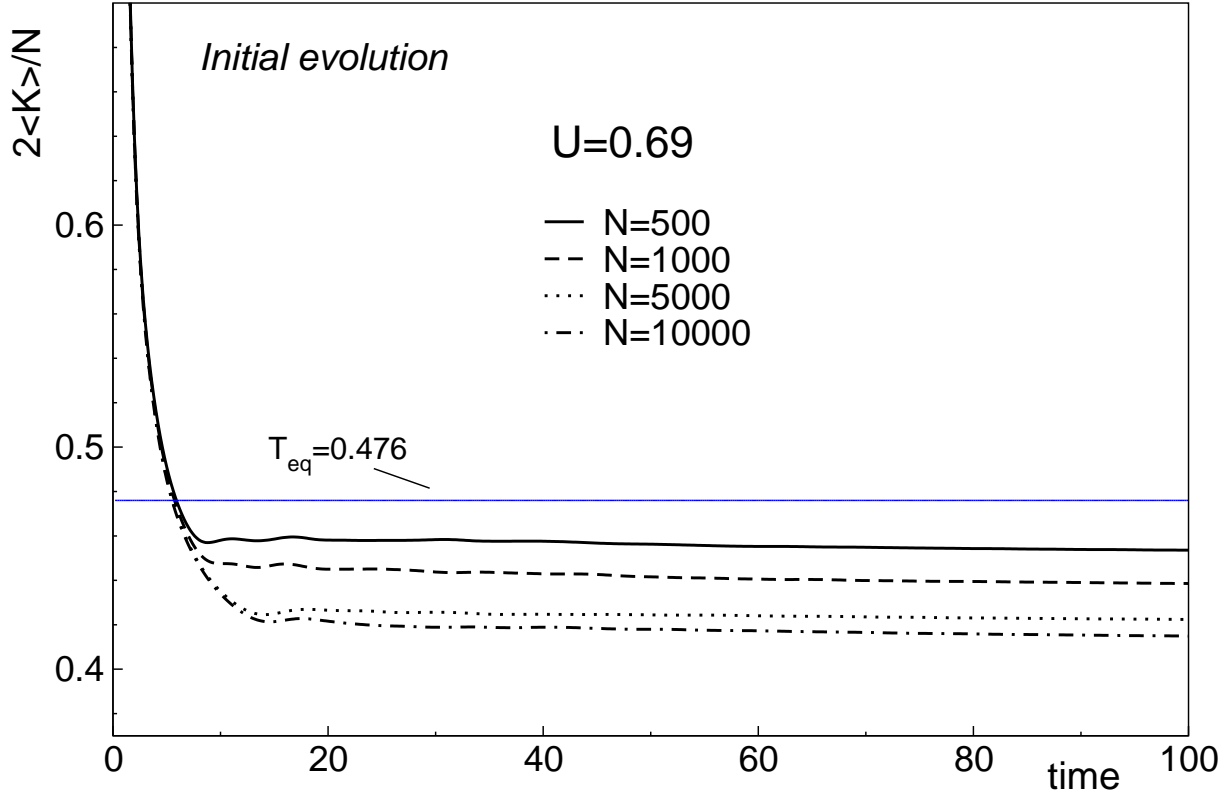,width=13truecm,angle=-90}
\end{center}
\caption{Initial time evolution of the quantity $2<K>/N$  for the
 HMF model. The  numerical simulations correspond to   different
 N sizes at $U=0.69$.
}
\end{figure}

In fig.2 we show the initial time evolution of $2<K>/N$  to display
 the fast collapse of the initial condition to the metastable
 regime. We start our simulations by considering the so-called
 \textit{water-bag initial conditions}
 , i.e. putting all the angles at zero and distributing the total
 energy uniformly over the momenta.  The figure shows that there
 is a rapid evolution, which is not size-dependent, towards the
 QSS state. In ref. [10],  we have also checked that  these simulations
 are not affected by the numerical accuracy of the integration
 scheme used. This is certainly true in the range  explored,
 with a relative error $10^{-7}  < \Delta E/E   <  10^{-3}$, and  
 demonstrates  the robustness
 of   these metastable  QSS against small perturbations .   
At this point, it is interesting to calculate the specific heat
  in correspondence of this slow relaxation in the energy region
 below the critical point.  The specific heat can be calculated
 from the fluctuations of kinetic energy, by  using the microcanonical
 formulas derived in the 60's by Lebowitz, Percus and Verlet
 (LPV formula) [18], i.e.
\begin{equation}
C_V={1 \over 2} \left[ 1 -2   \left(  {\Sigma \over T} \right)^2 \right]^{-1}
\end{equation}

\noindent
T being   the microcanonical temperature and $\Sigma$ the kinetic
 energy microcanonical fluctuation. A second alternative is another
 formula which was derived more recently by  Pearson, Haliciouglu
 and Tiller (PHT formula) [19]. The latter should be more precise,
 because it takes  into account finite-size effects and   is
 given by 
\begin{equation}
C_V= \left[ 2<K> <K>^{-1} + N (1-<K> <K>^{-1})   \right]^{-1}.
\end{equation}

\begin{figure}
\begin{center}
\epsfig{figure=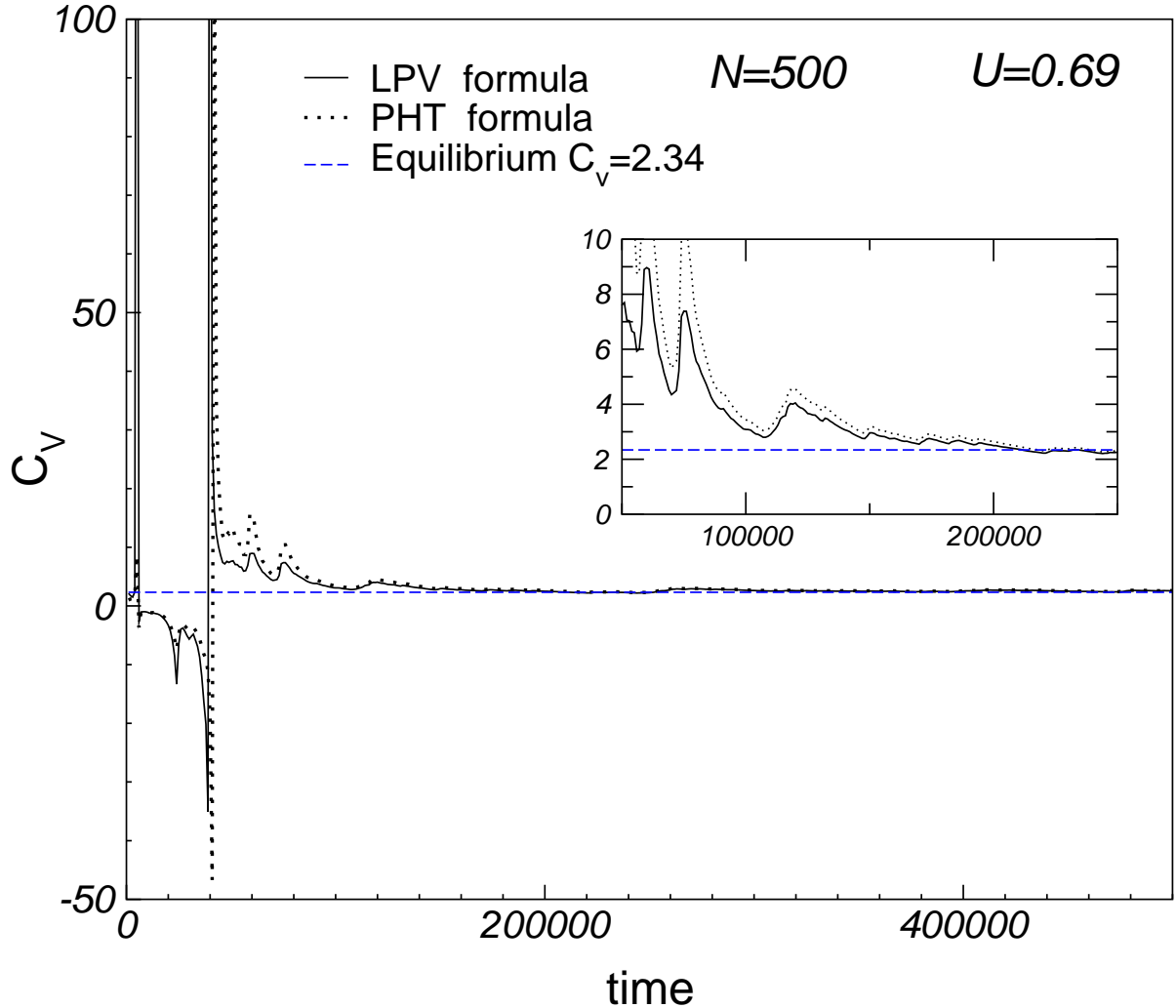,width=15truecm,angle=-90}
\end{center}
\caption{  Specific heat calculations for the HMF model as a
 function of time  in the case  N=500 and U=0.69. The two different
 formulas discussed in the text, LPV formula (4) and PHT formula
 (5), are compared in the plot  and in the inset magnification.
 The figure shows a  very similar time evolution and a nice convergence
 for long times  to the correct positive equilibrium value, indicated
 by the dashed line.
}
\end{figure}

\noindent
We report in fig.3 the specific heat calculated according to
 the eqs. (4) and (5) for N=500 and U=0.69. The figure shows
 a very similar time evolution for both formulas:   after some
 oscillations, in which the specific heat is negative, both numerical
 simulations converge towards the same correct equilibrium positive
 value [16], also indicated as a dashed line. 
Let us now focus our attention on the transient regime, where
 the specific heat is negative. We have found that, in the transient
 QSS regime, the system does not show a Gaussian velocity probability
 distribution, while for longer integration times, when the system
 finally relaxes

\begin{figure}
\begin{center}
\epsfig{figure=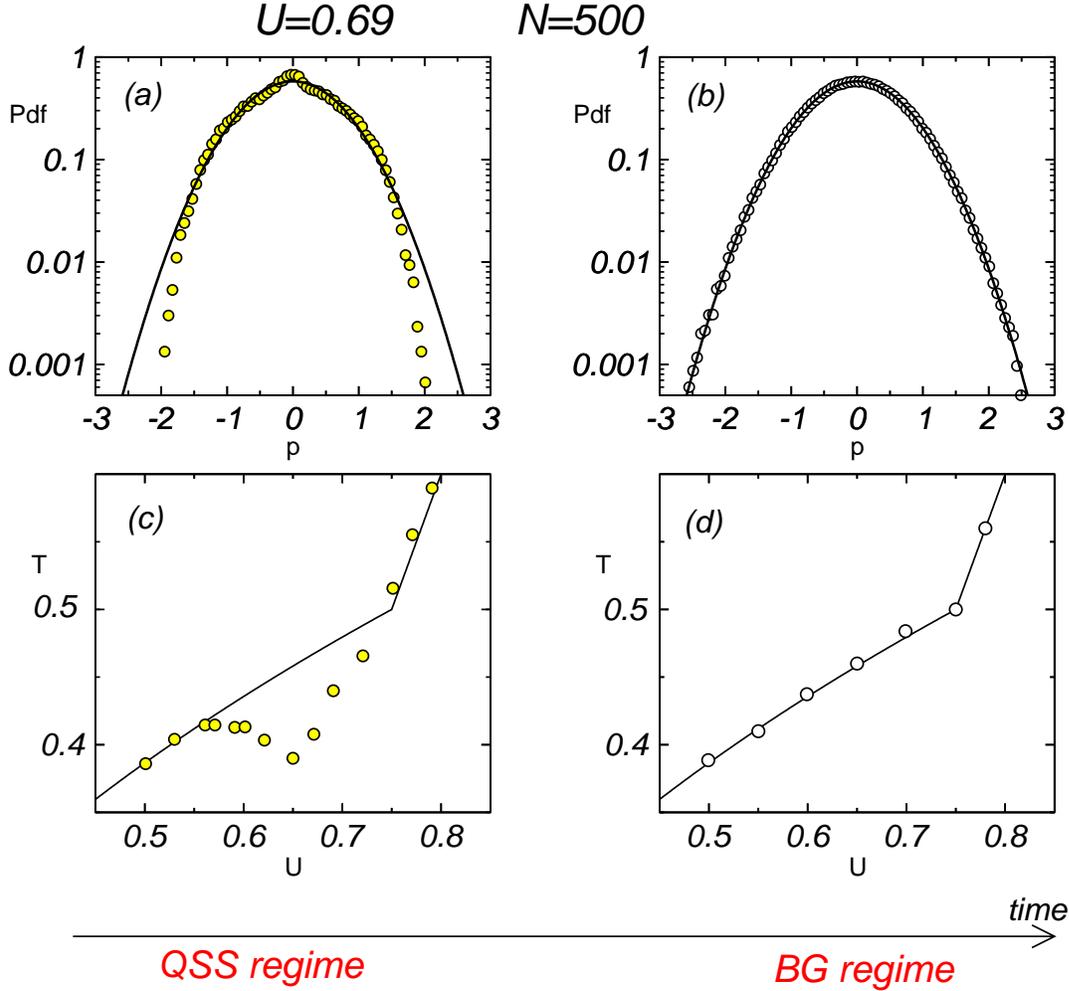,width=15truecm,angle=-90}
\end{center}
\caption{ Caloric curves and velocity pdfs  for the HMF model
  in the QSS regime and in the BG equilibrium regime  for N=500.}
\end{figure}

\noindent
to the equilibrium temperature, the velocity probability distribution
 (pdf) is perfectly Gaussian [10]. This result is nicely illustrated
 in fig.4. There   we plot   the numerical simulations for the
 caloric curve and the velocity pdfs both in the transient case,
 yellow circles in panels (c) and (a),  and in the equilibrium
 case, white circles in panels (d) and (b). A back-bending of
 the HMF caloric curve in the transient regime clearly coincides
 with a  non-Gaussian shape of the velocity pdf. In particular
 pdf tails are missing  and slow decaying velocity  correlations
 are present.
One can also show that the velocity pdfs are frozen in this anomalous
 distribution for all the duration of the metastable regime.
 This result can be easily understood  since  the force, which
 each spin feels, is almost zero in the QSS regime. In fact 
 the force on the single spin is given by the formula  
$F_i= -M_x sin \theta_i + M_y cos \theta_i$,
 where $M_x$ and  $M_y$  are the components of the magnetization vector.
 Thus being $M^2=0$  in the QSS regime, it follows  that  also $F_i=0$ .
 The system is attracted by the QSS state and then remains in
 a frozen state. Why it then relaxes to the BG equilibrium? The
 answer is simple. If the system size is finite, the magnetization
 is not exactly zero, a small noise ($O\sim {1 \over \sqrt{N}}$)
 always exists and this
  is  responsible for the final relaxation to the BG equilibrium
 regime.  The bigger the size, the smaller is the noise  and
 thus the longer is the lifetime of the QSS state as numerically
 found [10]. This fact implies the interesting result that, if
 one inverts the order  of the limits, i.e. takes first  the
 infinite size limit instead of the infinite time limit, the
 noise is perfectly zero and the system remains trapped in the
 QSS state for ever.  In [10] we have shown that the formalism
 proposed by Tsallis [8,9] seems to explain the shape of the
 velocity pdfs.

\begin{center}
\textbf{
4  CONCLUSIONS
}

\end{center}

Summarizing,we have shown an example of a nonextensive system where 
a negative specific heat is found in correspondence to quasi-stationary
 states (QSS) and  non-Gaussian velocity pdfs. This result which
 has been recently  confirmed   also  in  other    long-range
 interacting  models such as self-gravitating systems [7,12]
 and modified Lennard-Jones potentials [13] is due to the nonextensivity
 nature of the system into exam [14].  This fact is  a serious
 warning for a straightforward claiming of a standard equilibrium
 first-order phase transition in nuclear fragmenting systems.
 Although some sort of equilibration seems to be reached in multifragmentation,
  it is not certain whether  this corresponds to a complete relaxation.
 We stress the fact that  the temperature deviation from its
 equilibrium value, we have discussed   so far, is only of the
 order of 10\%  !  In this respect more detailed investigations
 should be done in order to further clarify the general theoretical
 scenario for nonextensive systems. But also more  precise experimental
 data are  very welcome and could be extremely useful to  understand
 deeper the intriguing nature of phase transitions in finite
 systems, a field surely at the frontier of  modern statistical
 mechanics.

\begin{center}
\textbf{
REFERENCES
}

\end{center}

\begin{flushleft}
[1] D.H.E. Gross, \textit{Microcanonical thermodynamics: phase transitions in small systems}
, Lecture Notes 
in Physics, Springer-Verlag Heidelberg 2001 and
 refs therein. See also the proceedings of this workshop.

[2]  P. Chomaz and F. Gulminelli, Nucl. Phys. A 647 (1999) 153

[3]	M.D'Agostino et al, Phys. Lett. B A473 (2000) 219.

[4]   M. Schmidt et al. Phys. Rev. Lett. 86 (2001) 1191.

[5]  K. Huang, \textit{Statistical mechanics}
, Wiley  (1987).

[6]  J. Barr\`{e}, D. Mukamel, S. Ruffo, Phys. Rev. Lett. 87
 (2001) 030601.

[7]  A. Torcini, M. Antoni, Phys. Rev. E 89 (1999) 2746.

[8]   C. Tsallis, J.Stat. Phys. 52  (1988)  479.

[9]  For an updated review of this generalized statistics see
 the proceedings of the conference NEXT2001 published in  Physica
 A 305 (2002).  An updated reference list is also available at
  http://tsallis.cat.cbpf.br/biblio.htm

 [10]V.Latora and A. Rapisarda,  Nucl. Phys. A   681 (2001) 406c;
 V. Latora, A. Rapisarda and C. Tsallis, Phys. Rev. E   64 (2001)
  056134 and Physica  A 305 (2002) 129.

[11] C.Tsallis, B.J.C. Cabral, A. Rapisarda and V. Latora, [cond-mat/0112266]
 submitted to Phys. Rev. Lett.

[12] Sota et al,  Phys. Rev. E  64 (2001) 05613; A. Taruya, M.
 Sakagami, Physica A (2002) in press [cond-mat/0107494].

[13] E.P. Borges and C. Tsallis, Physica A 305 (2002) 148.

[14] A. Campa,  A. Giansanti, D. Moroni, Physica A 305 (2002)
 137.

[15]	M. Antoni and S. Ruffo Phys. Rev. E 52  (1995) 2361.

[16]	V. Latora, A. Rapisarda and S. Ruffo, Phys. Rev. Lett. 
 80 (1998) 698, Physica D 131 (1999) 38 and  Phys. Rev. Lett.
 83 (1999) 2104.

[17] C. Anteneodo and C. Tsallis, Phys. Rev. Lett. 80 (1998)5313;
 A. Campa, A. Giansanti, D. Moroni and C.Tsallis, Phys. Lett.
 A 286 (2001) 251.

[18]  J.L. Lebowitz, J.K. Percus and L. Verlet, Phys. Rev. A
 153 (1967) 250.

[19] E.M. Pearson, T. Halicioglu and W.A. Tiller Phys. Rev. A
 32 (1985) 3030.

\end{flushleft}

\end{document}